\documentclass[11pt,a4paper]{article}
\usepackage{amsmath,amssymb}
\usepackage{graphicx}
\def\DR{\ensuremath{\overline{\mathrm{DR}}}}
\def\MS{\ensuremath{\overline{\mathrm{MS}}}}

\begin{document}

\title{\bf Some two-loop threshold corrections and three-loop renormalization group 
analysis of the MSSM}
\author{A.~V.~Bednyakov\footnote{{\bf e-mail}: bednya@theor.jinr.ru}
\\
\small{\em Joint Institute for Nuclear Research, 141980, Dubna, Russia}
}
\date{}
\maketitle

\begin{abstract}
Two-loop threshold corrections for both the strong coupling constant 
and Yukawa couplings of heavy SM fermions are
considered in the context of Minimal Supersymmetric Standard Model (MSSM).
With the help of the well-known SOFTSUSY code the dependence of the corrections on universal MSSM parameters 
is analyzed. 
For a consistent study of the influence of the corrections on SUSY mass spectrum three-loop renormalization
group equations are implemented. A particular scenario (SPS4) is considered and 
the shifts of the particle masses due to contributions of different threshold corrections are presented. 
Additionally, the impact on certain forbidden regions of the parameter space is studied.
\end{abstract}
\section{Introduction}
Minimal Supersymmtric Standard Model possesses many remarkable features that allows one to 
think of it as of the most viable candidate for a theory that describes physics beyond the Standard Model. 
Unfortunately, it has a lot of parameters which are related to unknown masses of SUSY particles. 
One way to deal with the problem is to use universal parameters at some high-energy scale  
and extrapolate them to low energies with the help of renormalization group equations (RGEs). 
Most of the computer codes \cite{Allanach:2003jw} used to obtain the SUSY 
spectrum %and the values of running parameters at some particular scale 
incorporate two-loop RGEs together with one-loop threshold corrections.
The latter allow one to calculate boundary values for gauge and Yukawa couplings from the given low-energy input 
in a consistent way. 

The necessity of threshold corrections is tightly related to the fact that the theories defined in a
minimal subtraction scheme cease to satisfy the Appelquist-Carazzone theorem \cite{Appelquist:1974tg} in a naive form. 
Heavy degrees of freedom
contribute to low energy observables even in the limit when the corresponding masses tend to infinity introducing 
potentially large logarithmic corrections. However,
the latter are local and can be absorbed in the definition of the couplings of the effective theory and summed up with
the help of a renormalization group.
There exists a so-called matching (or decoupling) procedure that allows one to routinely calculate threshold corrections.

One-loop decoupling corrections can be found in~\cite{Pierce:1996zz}. It turns out that sometimes
they can significantly change (e.g. by 40~\% for the $b$-quark mass) 
the value of the parameters defined in the MSSM with respect to that in the SM.  
In such cases it is reasonable to calculate contributions from the next order of perturbation theory (PT). 
 
It took some time to find leading two-loop corrections to matching relations between 
the strongest SM and MSSM couplings. Strong coupling was considered in~\cite{Harlander:2005wm}. 
The corresponding results for heavy SM fermion masses were found in~\cite{Bednyakov:2002sf,Bednyakov:2004gr} 

It is worth mentioning that the low-energy input to the MSSM can be given in terms of running SM parameters 
at the electroweak scale. However, there exists a distinction in minimal renormalization schemes usually 
used in the SM and MSSM. It comes from the fact that supersymmetry  requires 
vector bosons to be accompanied by so-called $\varepsilon$-scalars. The latter maintain the balance between fermionic and 
bosonic degrees of freedom in a dimensionally regularized theory. As a consequence, most calculations in the context
of the MSSM made use of the so-called \DR-scheme \cite{Siegel:1979wq} with $\varepsilon$-scalars implicitly (or explicitly) 
taken into account.  
This fact leads to the problem of $\DR\to \MS$ conversion since for the Standard Model the $\MS$ scheme is more suitable.  
The corresponding parameter redefinition can be done in a conventional way by comparison of certain (invariant) quantities
calculated in both schemes within the same model (SM or MSSM). However, this route leads to appearance of 
unphysical ``evanescent'' couplings if such relations are considered in the SM \cite{Jack:1993ws} or  
to supersymmetry breaking at rigid level in the MSSM. 
Recently, it was proposed that $\DR$ parameters of the MSSM and $\MS$ parameters of the SM can be directly related
since unphysical $\varepsilon$-scalars can be treated along the same lines as heavy degrees of freedom  
during calculation of decoupling corrections \cite{Bednyakov:2007vm}. 

\section{Two-loop matching}
Let me introduce necessary  notation and briefly discuss practical prescription for the matching procedure 
\cite{Chetyrkin:2000yt}.
The task is to find relations of the type
\begin{equation}
\underline{A}(Q)= \zeta_A(\bar\mu) \cdot A(Q)
\label{dec_rel}
\end{equation}
where $\underline{A}(Q)$ and $A(Q)$ are some running parameters (e.g., $\alpha_s$) defined at the 
renormalization scale $Q$ in the effective (SM) and the 
fundamental (MSSM) theories, respectively. 
The quantity 
\begin{equation}
\zeta_A(Q) = 1 + \delta \zeta^{(1)}_A(Q) + \delta\zeta^{(2)}(Q)\cdots
\end{equation}
is called ``decoupling constant'' for the parameter $A$ and can be calculated order-by-order
in PT. The relation \eqref{dec_rel} allows one to express $A$ in terms of $\underline A$ and other parameters of the fundamental theory at any scale $Q$. In practice, however, the scale is chosen in a way to minimize  
uncertainties due high orders of PT.

A convenient way to find the expression for $\zeta_A(Q)$ is to consider bare quantities 
and find ``bare'' decoupling constant $\zeta_{A,0}$ 
\begin{equation}
\underline{A}_0 = \zeta_{A,0} \cdot A_0
\label{bare_dec_rel}
\end{equation}
by demanding that bare Green functions calculated in both theories coincide\footnote{
	Field redefinition is also required in this case.} 
in the limit of vanishing external momenta and masses of the considered effective field theory. 
In this case all the diagrams without heavy degrees of freedom % that have mass denoted here by $M$ 
also vanish and the problem
is reduced to calculation of bubble integrals with at least one heavy line with mass denoted here by $M$. 
Given \eqref{bare_dec_rel} one expresses
bare quantities in terms of renormalized ones 
\begin{eqnarray}
\underline{A_0} & = & \underline{Z_A} \, (\underline{A}) \,
	\underline{A}, \qquad
A_0  =  Z_A \, (A, B) \, A, \nonumber\\
	\zeta_{A}
& = & 
\Bigl[ Z_A(A,\, B) \Bigr]
\Bigl[\,\underline{Z_A}\,(\underline{A}) \,\Bigr]^{-1}
	\,
	\zeta_{A,0}(Z_A \, A, Z_B \, B, \, Z_M M)
	. \label{decoupling:renormalized}
\end{eqnarray}
and after a proper re-expansion  obtains \eqref{dec_rel}. In \eqref{decoupling:renormalized} $B$
stands for couplings presented in the fundamental theory that are absent in the effective one. 

Let us consider one- and two-loop contributions to the decoupling constants for the strong coupling, 
$\zeta_{\alpha_s}$, the mass of bottom-quark, $\zeta_{m_b}$, and the tau-lepton mass, $\zeta_{m_{\tau}}$. 
Due to the fact that full two-loop calculations require evaluation of many thousands of diagrams, 
it is reasonable to consider a simplified setup and neglect electroweak gauge couplings. 
In this case, five-flavor QCD with free tau-lepton plays the role of effective theory instead of the SM.
The corresponding limit of the MSSM (``gauge-less'' limit) is treated as a more fundamental theory and 
is used to calculate threshold corrections.

\begin{figure}[tb]
\begin{center}
\begin{tabular}{ll}
 \hspace{2.6cm}{\small $\delta \zeta^{(1)}_{m_b}(M_Z)$, \%} & 
 \hspace{2.6cm}{\small $\delta \zeta^{(2)}_{m_b}(M_Z)$, \%}  \\
\includegraphics[scale=1]{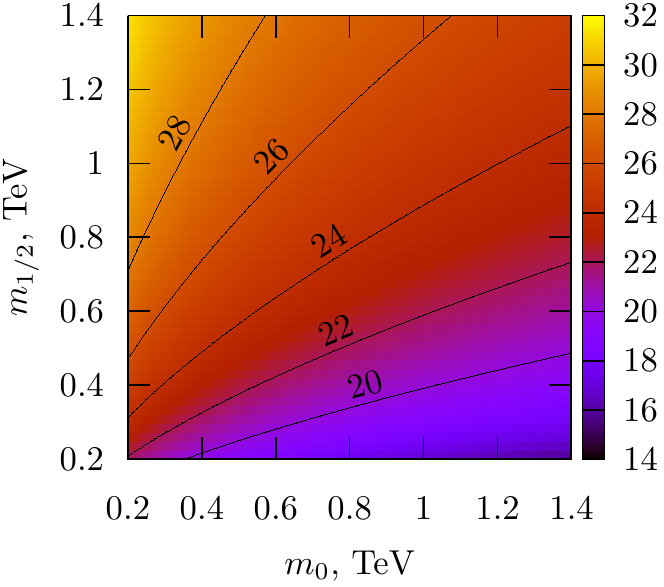} & 
\includegraphics[scale=1]{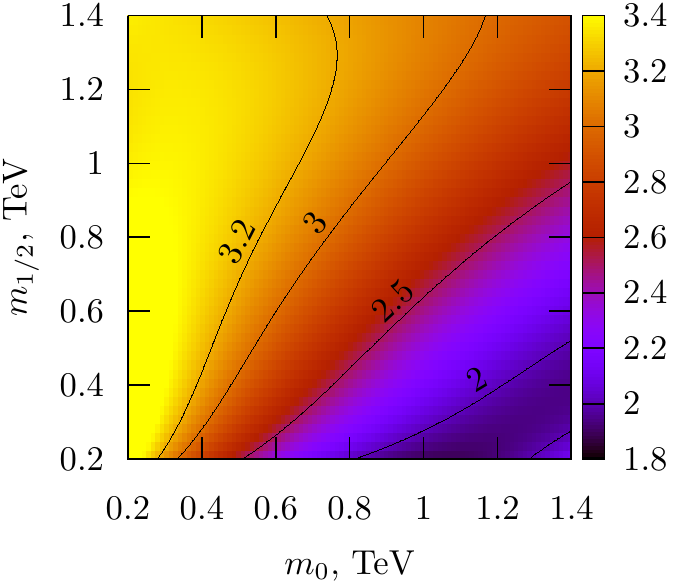} \\
\hspace{2.6cm}{\small $\Delta^{(1)} m_t/m_t (M_Z)$, \%} & 
 \hspace{2.6cm}{\small $\Delta^{(2)} m_t/m_t (M_Z)$, \%}  \\
\includegraphics[scale=1]{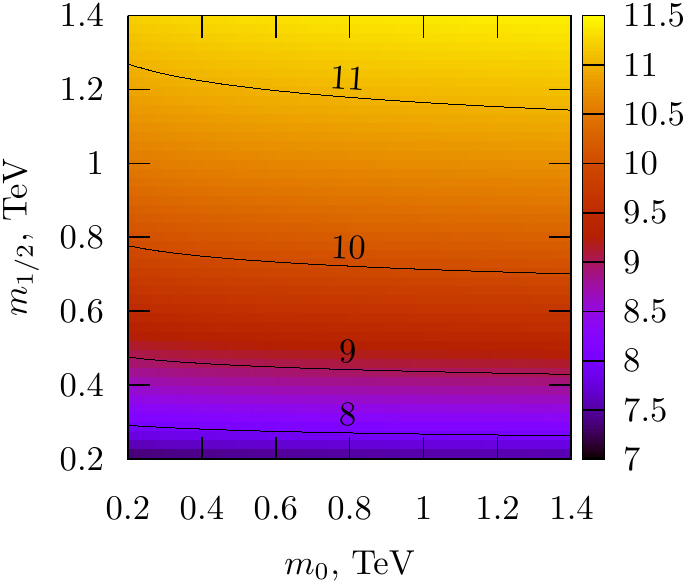} & 
\includegraphics[scale=1]{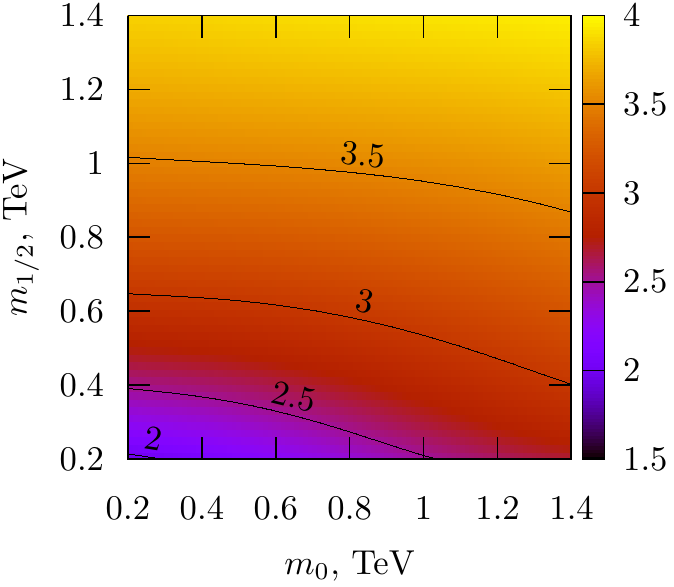} 
\end{tabular}
\end{center}
\caption{One- and two-loop corrections to the $b$-quark decoupling constant $\zeta_{m_b}$ and
	the pole mass $M_t$ of $t$-quark as functions of universal parameters $m_0$ and $m_{1/2}$ 
	for $\tan\beta=50$ and $A_0=0$.}
\label{fig:1}
\end{figure}

It is worth mentioning that since the $t$-quark mass is of the order of electroweak scale, it is reasonable 
to use the pole mass $M_t$ to find the boundary condition for corresponding running parameter $m^{\DR}_t(\bar\mu)$
in the MSSM
\begin{equation*}
	M_t = m^{\DR}_t(\bar\mu) \left(1 + \frac{\Delta^{(1)} m_t}{m_t} + \frac{\Delta^{(2)} m_t}{m_t} + \cdots \right)
\end{equation*}
	Here $\Delta^{(l)} m_t/m_t$ stands for $l$-loop contribution to the relation between $M_t$ and $m^{\DR}$.

Since two-loop expressions are very lengthy, it is convenient to analyze them numerically. Figure~\ref{fig:1} 
shows the dependence of quark mass corrections on the MSSM universal parameters $m_0$ and $m_{1/2}$.
Large value of $\tan\beta=50$ has been chosen since in this case loop corrections to the bottom-quark (and tau-lepton) mass 
are significantly enhanced. For the $t$-quark, contribution due to Yukawa couplings is suppressed 
by inverse $\tan\beta$ and is not taken into account.   

I would like to stress that small values of $\delta^{(2)}_{m_b}$ result from 
large cancellations of contributions proportional to strong and Yukawa couplings. % $\alpha_y \equiv y^2/(4\pi)$.
For example, in the region of small $m_0$ and large $m_{1/2}$ the two-loop $\mathcal{O}(\alpha_s^2)$ contribution
can reach 10~\%. However, negative  corrections due to Yukawa couplings 
lower this value down to a few per cent only.

\begin{figure}[tb]
\begin{center}
\begin{tabular}{ll}
 \hspace{2.6cm}{\small $\delta \zeta^{(1)}_{m_\tau}(M_Z)$, \%} & 
 \hspace{2.6cm}{\small $\delta \zeta^{(2)}_{m_\tau}(M_Z)$, \%}  \\
\includegraphics[scale=1]{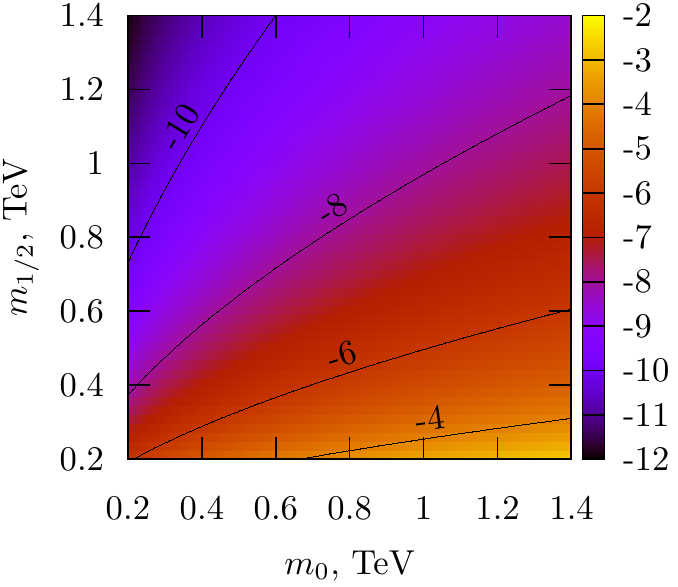} & 
\includegraphics[scale=1]{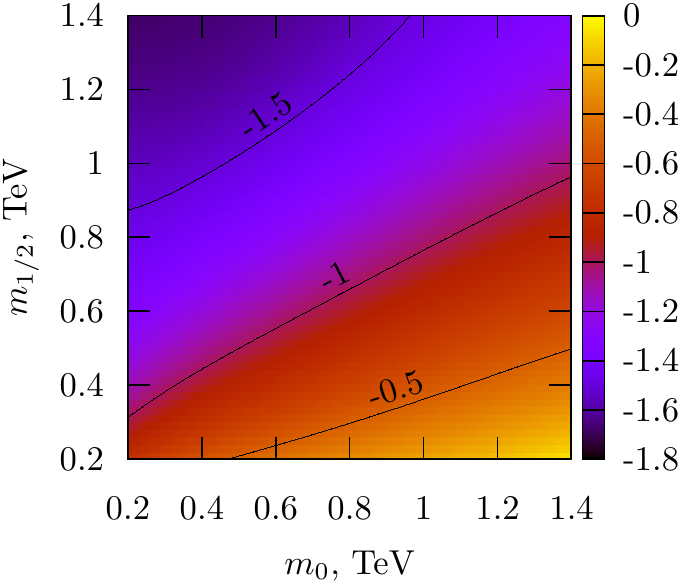}  \\
\hspace{2.6cm}{\small $\delta \zeta^{(1)}_{\alpha_s} (M_Z)$, \%} & 
 \hspace{2.6cm}{\small $\delta \zeta^{(2)}_{\alpha_s} (M_Z)$, \%}  \\
\includegraphics[scale=1]{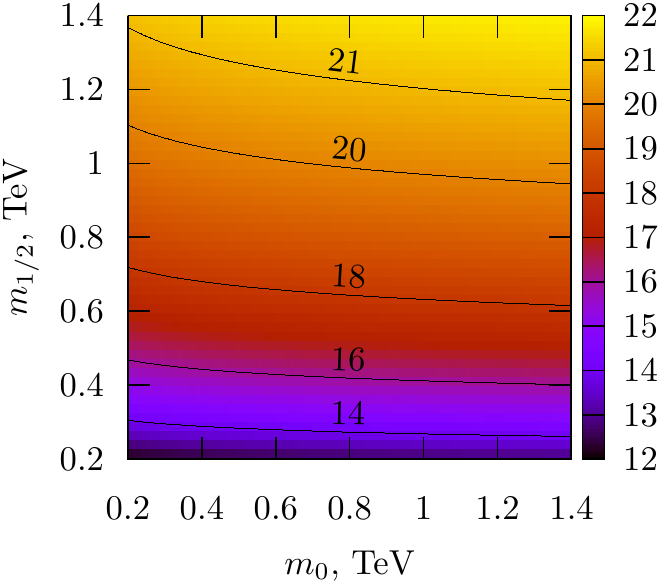} & 
\includegraphics[scale=1]{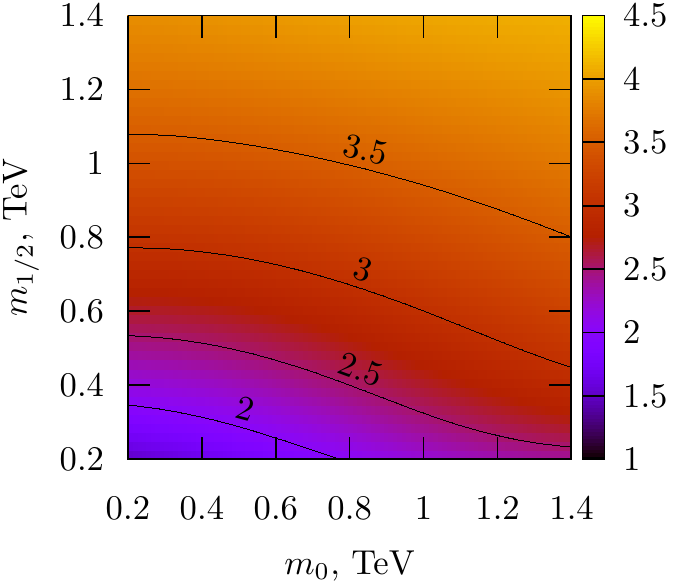} 
\end{tabular}
\end{center}
\caption{The dependence of threshold corrections to $\tau$-lepton mass (upper row) and strong coupling $\alpha_s$
	(lower row) on $m_0$ and $m_{1/2}$. Large value of $\tan\beta=50$ is assumed, and $A_0 = 0$. }
\label{fig:2}
\end{figure}

In the case of $\tau$-lepton (see upper row in Fig.~\ref{fig:2}) there are no corrections due to strong 
interactions and the net effect is negative. The two-loop corrections are naturally small but 
exceed the uncertainty in the experimental value $M_\tau=1776.84(17)$ MeV.

Finally, two-loop decoupling corrections for the strong coupling $\alpha_s$ were obtained in~\cite{Harlander:2005wm}. 
Numerical values can be found in Fig.~\ref{fig:2} (lower row). The corrections are small in comparison with one-loop result. 
Nevertheless, they can be important in studying gauge constant unification.  

It is interesting to note that the two-loop corrections for $\alpha_s$, $m_b$ and $m_t$ 
can be approximated by the same value of 2-3~\% for the whole considered region\footnote{The situation changes 
only slightly with variation of $\tan \beta$.} in  the $m_0$--$m_{1/2}$ plane.

All the above-mentioned corrections were taken into account in the  
three-loop renormalization group analysis of the MSSM that will be presented in the next section.  

\section{Three-loop analysis of the MSSM}

It is known that for a self-consistent $L$-loop study of the MSSM one needs to known threshold corrections up to $(L-1)$ 
order. Three-loop beta-functions for a rigid (supersymmetric) part of the MSSM Lagrangian were calculated by means  
of superfield Feynman diagram technique \cite{Ferreira:1996ug}. With the help of the spurion formalism developed 
in \cite{spurion} beta-functions for soft supersymmetry-breaking terms can be easily found from the results of 
\cite{Ferreira:1996ug}. 

\begin{figure}[thb]
\begin{center}
\begin{tabular}{cc}
\includegraphics[scale=0.55]{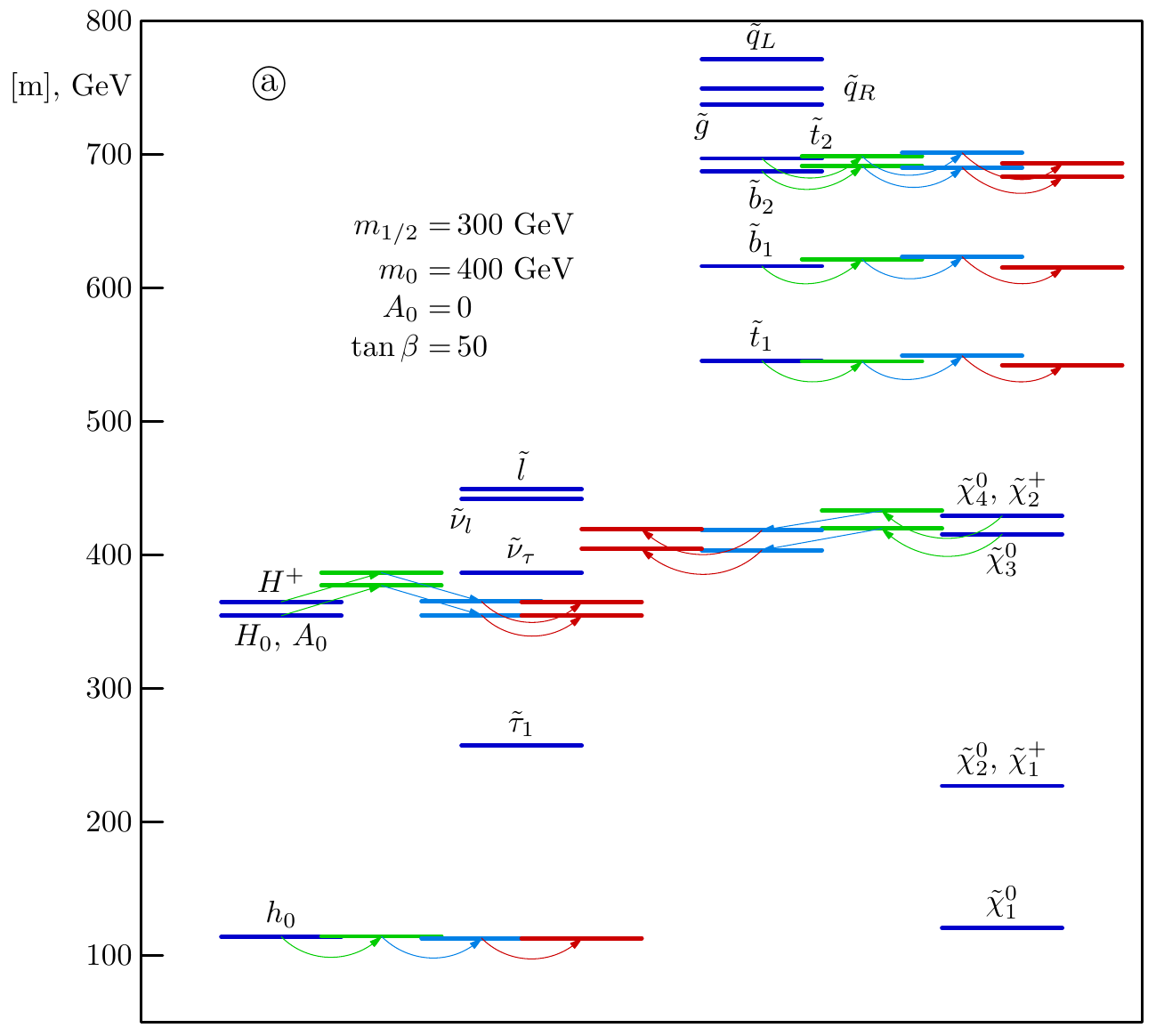} &
\includegraphics[scale=0.55]{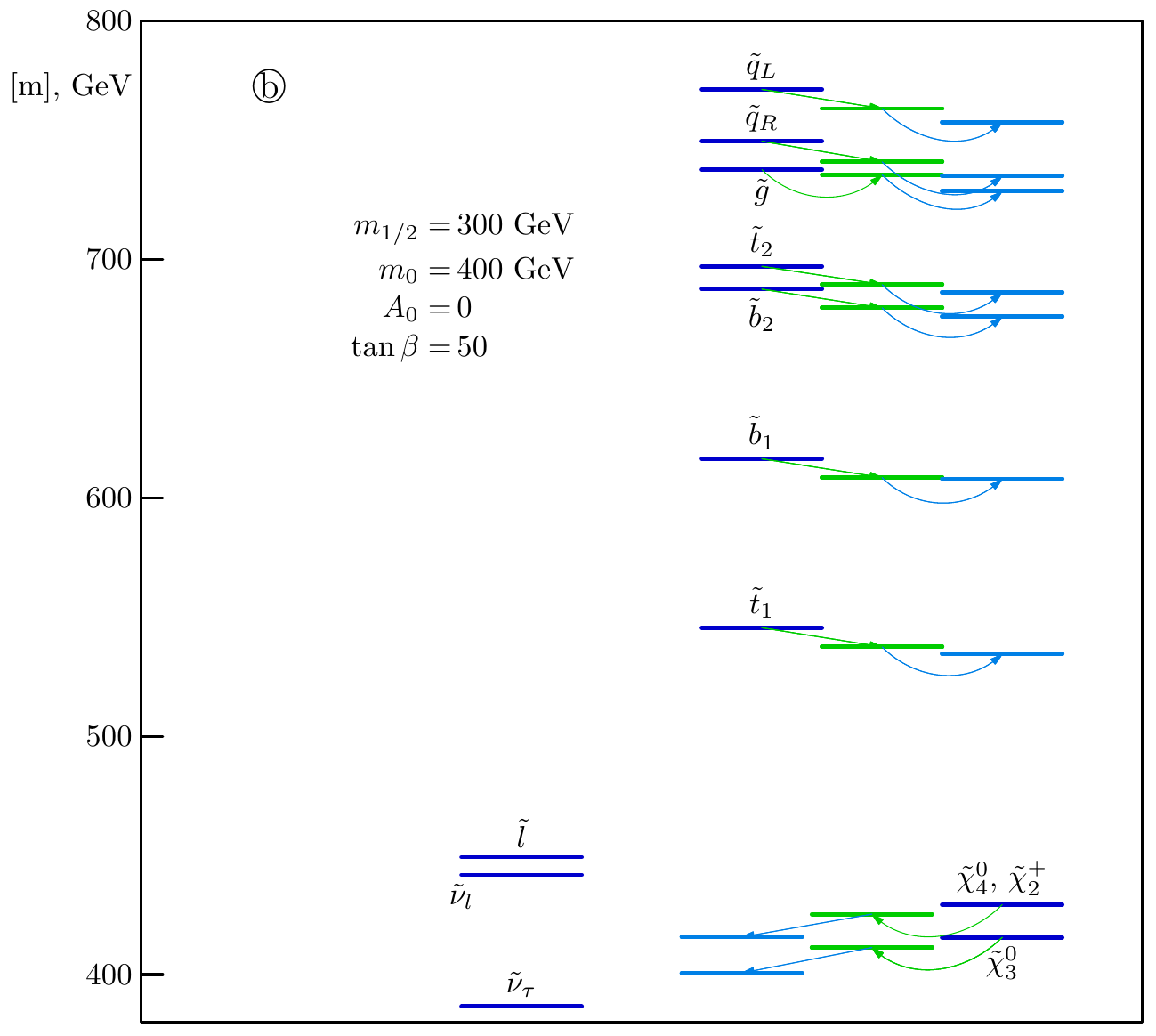} 
\end{tabular}
\end{center}
\caption{Initial spectrum for SPS4 scenario obtained with original SOFTSUSY code is shown together with shifts 
due to calculated two-loop decoupling corrections (a) 
and three-loop terms in RGEs (b). Left plot (a) corresponds to subsequent addition of $\delta \zeta_{m_b}$ (green), 
$\Delta m_t/m_t$ (blue), and $\delta \zeta_{\alpha_s}$ (red) into the code without modification of RGEs. 
Right plot (b) demonstrates the impact of three-loop RGEs alone (green arrows) and the subsequent inclusion of 
all the threshold corrections (blue). Corrections that are less than~1\% at all the steps are not shown.}
\label{fig:spec}
\end{figure}

The first three-loop analysis
of the MSSM was performed in \cite{Jack:2004ch} and it was found that the effect of three-loop running on the SUSY spectrum is small for weakly interacting particles but is larger for squark masses (1-5~\%). However, two-loop threshold corrections
were not available at those times. 
In this talk I will present the results of a more consistent study based on modified version of the SOFTSUSY \cite{Allanach:2001kg} code\footnote{Available from the author.} which takes into account both three-loop RGEs and two-loop decoupling corrections discussed earlier. 

As a benchmark scenario the so-called SPS4 point \cite{Allanach:2002nj} was chosen. For this point  
one has $m_0 = 400$~GeV, $m_{1/2} = 300$~GeV, $A_0 = 0$ at the GUT scale, 
and  $\tan\beta = 50$ at the electroweak scale. 
In order to visualize the impact of additional two-loop terms in
decoupling corrections and three-loop contributions to beta-functions, the 
spectrum produced by the SOFTSUSY code is presented for four cases. 

In Fig.~\ref{fig:spec}a one sees how two-loop corrections to  $b$- (green arrows) and $t$-quark (blue arrows) 
masses modify the initial spectrum of SPS4 obtained with the original SOFTSUSY 3.1 code.
Quark Yukawa couplings influence significantly the running of soft masses for the corresponding Higgs bosons. 
This, in turn, leads to the relatively large shifts  in the masses of heavy higgses (4-5~\%) and higgsinos  (2-3~\%). 
However, the overall result for heavy higgs masses is small (less than 1~\%) due to cancellations.
Inclusion of $\delta^{(2)}\zeta_{\alpha_s}$ (red arrows) slightly lowers the masses of third generation 
squarks below the initial value.

This picture for quark masses and strong coupling was obtained without the inclusion of three-loop terms in RGEs.
Figure~\ref{fig:spec}b shows the impact of these additional terms. First of all, for comparison with the results of Ref.~\cite{Jack:2004ch} the spectrum was calculated without two-loop threshold corrections (green arrows). 
As it was noticed by the previous authors the corrections due to three-loop terms are small 
and mostly influence strongly interacting particles (by 1-2~\%). 
After the inclusion of calculated two-loop decoupling corrections one can see additional shifts 
for squarks, gluino, and neutralino/chargino with large higgsino component.
In the end, one has 1-2~\% overall correction to the masses of strongly interacting particles 
and 3~\% correction to the higgsino masses.  It is also interesting to note that the inclusion of three-loop RGEs lowers
the value of the lightest higgs boson mass from 114 GeV down to 113 GeV (not shown in Fig.~\ref{fig:spec}b).

\section{Summary and Conclusion}
	Some two-loop threshold corrections to the SM parameters were calculated and numerically studied
in a wide region of the parameters space of the MSSM. It turns out that for the 
considered region two-loop contributions to the 
decoupling constants of the strong coupling and heavy quark masses are of the order of 2-4~\% and do not depend significantly
on $\tan \beta$. Since tau-lepton does not participate in strong interactions, the corresponding correction to its mass
is smaller than that of quarks. Nevertheless, for large $\tan\beta$ it can reach the value of 1~\% which exceeds current
experimental uncertainty of the tau-lepton pole mass.

	A proper way to use the above-mentioned two-loop quantities is to incorporate 
them together with three-loop RGEs in a code used to calculate the SUSY mass spectrum. For this purpose the SOFTSUSY 
package has been modified and it was found how the corrections influence the spectrum. 
A numerical study was performed for a particular scenario SPS4 with large $\tan\beta$ for which certain 
decoupling corrections are expected to be large. The overall effect of the three-loop running on the mass spectrum
turns out to be small and does not exceed a few per cent. In comparison with previous studies it was found 
that the inclusion of the decoupling corrections besides lowering squark masses leads to a decrease in 
gluino and higgsino masses. 

\begin{figure}[h]
\begin{center}
\includegraphics{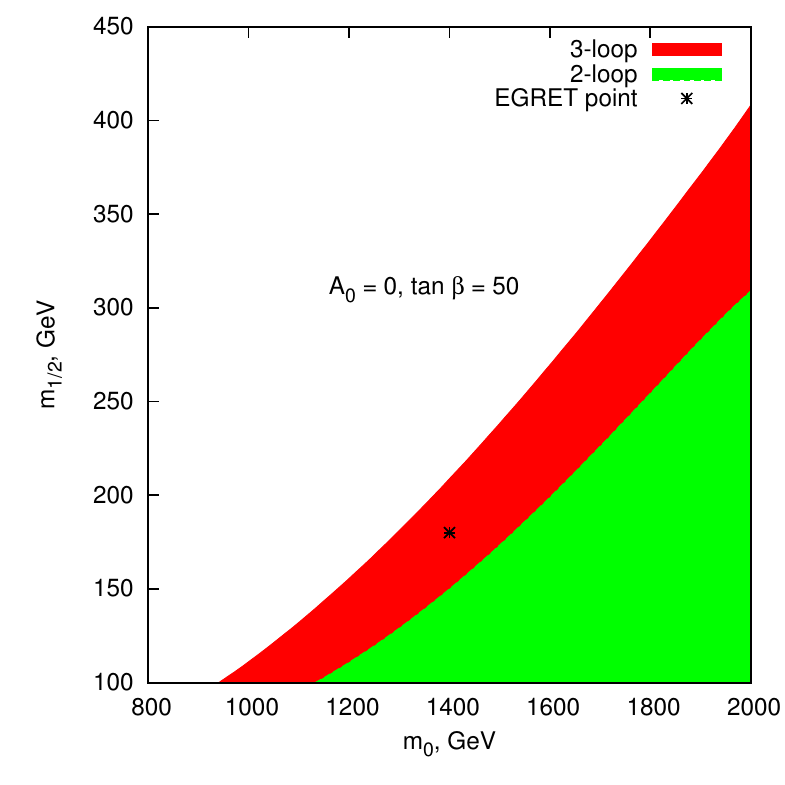}
\end{center}
\caption{Forbidden regions in the $m_0-m_{1/2}$ plane where no EWSB occurs. 
The green region was obtained with the help of two-loop RGEs and the red one --- with three-loop terms.
It is clear that the so-called EGRET \cite{deBoer:2005bd} point is not allowed by three-loop analysis.}
\label{fig:noEWSB}
\end{figure}

It is also interesting to study the influence 
on allowed regions in the parameter space. For example, the boundary that separates the regions with and without electroweak symmetry breaking (EWSB) can be significantly shifted. Figure~\ref{fig:noEWSB} shows how the forbidden region increases after taking into account three-loop RGEs and the corresponding threshold effects. The so-called ``EGRET point'' proposed in \cite{deBoer:2005bd} 
is also shown and it looks like three-loop evolution excludes it. 
Therefore, one should be careful when choosing
particular values of parameters near boundaries of the allowed region.

Moreover, although being small the calculated corrections give us an opportunity to estimate theoretical uncertainties of
the SUSY parameters fitted with the help of two-loop RGEs. 
This seems to be more reliable than (or at least complementary to) the comparison between the results 
of different computer codes. 

At the end, I would like to thank the organizers of the ``Quarks 2010'' seminar for the opportunity to participate in such a nice event. Financial support
from RFBR grant 08-02-00856-a is kindly acknowledged.
%Since quark mass corrections are positive at the electroweak scale which usually used for matching 
%corresponding running Yukawa couplings become smaller in MSSM and influence Higgs soft masses $m^2_{H_1}$ and $m^2_{H_2}$
%significantly. Minimization condition for Higgs potential gives us the relations
%\begin{eqnarray*}
%\mu^2 & = &   \frac{m_{H_1}^2- m_{H_2}^2 \tan^2 \beta}{\tan^2 \beta - 1} - \frac{1}{2} M_Z^2, \\
%M_A^2 & = & m_{H_1}^2 + m_{H_2}^2 + |\mu|^2 = \frac{m_{H_2}^2- m_{H_1}^2 }{\cos 2 \beta} - M_Z^2 
%\end{eqnarray*}  
%	where parameters $\mu$ and $M_A^2$ sets the masses of higgsino and heavy Higgses.    


\begin{thebibliography}{10}
\providecommand{\url}[1]{\texttt{#1}}
\providecommand{\urlprefix}{URL }
\expandafter\ifx\csname urlstyle\endcsname\relax
  \providecommand{\doi}[1]{doi:\discretionary{}{}{}#1}\else
  \providecommand{\doi}{doi:\discretionary{}{}{}\begingroup
  \urlstyle{rm}\Url}\fi
\providecommand{\eprint}[2][]{\url{#2}}

\bibitem{Allanach:2003jw}
B.~C. Allanach, S.~Kraml and W.~Porod,
\newblock JHEP {\bf 03}, 016 (2003),
\newblock hep-ph/0302102

\bibitem{Appelquist:1974tg}
T.~Appelquist and J.~Carazzone,
\newblock Phys. Rev. D {\bf 11}, 2856 (1975)

\bibitem{Pierce:1996zz}
D.~M. Pierce, J.~A. Bagger, K.~T. Matchev and R.-j. Zhang,
\newblock Nucl. Phys. B {\bf 491}, 3 (1997),
\newblock hep-ph/9606211

\bibitem{Harlander:2005wm}
R.~Harlander, L.~Mihaila and M.~Steinhauser,
\newblock Phys. Rev. D {\bf 72}, 095009 (2005),
\newblock hep-ph/0509048

\bibitem{Bednyakov:2002sf}
A.~Bednyakov, A.~Onishchenko, V.~Velizhanin and O.~Veretin,
\newblock Eur. Phys. J. C {\bf 29}, 87 (2003),
\newblock hep-ph/0210258

\bibitem{Bednyakov:2004gr}
A.~Bednyakov and A.~Sheplyakov,
\newblock Phys. Lett. B {\bf 604}, 91 (2004),
\newblock hep-ph/0410128

\bibitem{Siegel:1979wq}
W.~Siegel,
\newblock Phys. Lett. B {\bf 84}, 193 (1979)

\bibitem{Jack:1993ws}
I.~Jack, D.~R.~T. Jones and K.~L. Roberts,
\newblock Z. Phys. C {\bf 62}, 161 (1994),
\newblock hep-ph/9310301

\bibitem{Bednyakov:2007vm}
A.~V. Bednyakov,
\newblock Int. J. Mod. Phys. A {\bf 22}, 5245 (2007),
\newblock 0707.0650

\bibitem{Chetyrkin:2000yt}
K.~G. Chetyrkin, J.~H. Kuhn and M.~Steinhauser,
\newblock Comput. Phys. Commun. {\bf 133}, 43 (2000),
\newblock hep-ph/0004189

\bibitem{Ferreira:1996ug}
P.~M. Ferreira, I.~Jack and D.~R.~T. Jones,
\newblock Phys. Lett. B {\bf 387}, 80 (1996),
\newblock hep-ph/9605440

\bibitem{spurion}
I.~Jack and D.~Jones,
\newblock Phys. Lett. B {\bf 415}, 383 (1997),
\newblock hep-ph/9709364;~ 
L.~V.~Avdeev, D.~I.~Kazakov and I.~N.~Kondrashuk,
\newblock Nucl. Phys. B {\bf 510}, 289 (1998), 
\newblock hep-ph/9709397;~
I.~Jack, D.~R.~T.~Jones and A.~Pickering,
\newblock Phys. Lett. B {\bf 432}, 10 (1998),
\newblock hep-ph/9803405

\bibitem{Jack:2004ch}
I.~Jack, D.~R.~T. Jones and A.~F. Kord,
\newblock Ann. Phys. {\bf 316}, 213 (2005),
\newblock hep-ph/0408128

\bibitem{Allanach:2001kg}
B.~C. Allanach,
\newblock Comput. Phys. Commun. {\bf 143}, 305 (2002),
\newblock hep-ph/0104145

\bibitem{Allanach:2002nj}
B.~C. Allanach \emph{et~al.},
\newblock Eur. Phys. J. C {\bf 25}, 113 (2002),
\newblock hep-ph/0202233

\bibitem{deBoer:2005bd}
W.~de~Boer, C.~Sander, V.~Zhukov, A.~V. Gladyshev and D.~I. Kazakov,
\newblock Phys. Lett. B {\bf 636}, 13 (2006),
\newblock hep-ph/0511154

\end{thebibliography}
\end{document}